\documentclass[a4paper]{article}

\usepackage{INTERSPEECH2022}
\usepackage{amsmath}
\usepackage{caption}
\usepackage{subcaption}
\usepackage{wasysym}
\usepackage{marvosym}

\usepackage{url}
\usepackage[hidelinks]{hyperref}
\hypersetup{
    colorlinks=true,
    linkcolor=black,
    filecolor=black,
    urlcolor=black,
    citecolor=black,
    breaklinks=true
}

\def\genbox#1#2#3#4#5#6{
    \leavevmode\raise#4bp\hbox to#5bp{\vrule height#5bp depth0bp width0bp
    \pdfliteral{q .5 w \csname #2COLOR\endcsname\space RG
                       \csname #3PDF\endcsname{#5}{#6} S Q
             \ifx1#1 q \csname #2COLOR\endcsname\space rg 
                       \csname #3PDF\endcsname{#5}{#6} f Q\fi}\hss}}
\def\sqbox      #1#2{\genbox{#1}{#2}  {sq}       {0}   {4.5}  {2.25}}

\title{Generating gender-ambiguous voices for\\ privacy-preserving speech recognition}
\name{Dimitrios Stoidis, Andrea Cavallaro}
\address{Centre for Intelligent Sensing, Queen Mary University of London, UK}
\email{dimitrios.stoidis@qmul.ac.uk, a.cavallaro@qmul.ac.uk}

\begin{document}

\maketitle
\begin{abstract}
  Our voice encodes a uniquely identifiable pattern which can be used to infer private attributes, such as gender or identity, that an individual might wish not to reveal when using a speech recognition service. To prevent attribute inference attacks alongside speech recognition tasks, we present a generative adversarial network, GenGAN, that synthesises voices that conceal the gender or identity of a speaker. The proposed network includes a generator with a U-Net architecture that learns to fool a discriminator. We condition the generator only on gender information and use an adversarial loss between signal distortion and privacy preservation. We show that GenGAN improves the trade-off between privacy and utility compared to privacy-preserving representation learning methods that consider gender information as a sensitive attribute to protect.

\end{abstract}


\section{Introduction}\label{sec:intro}

The human voice is shaped by the physical characteristics of the speaker and the spoken language. The voiceprint, which uniquely defines each individual~\cite{voiceprint_identification}, contains attributes of the speaker that can be inferred by voice-based services~\cite{userprofiling, VC_adv}. To mitigate this risk and protect the identity of a speaker in Automatic Speech Recognition (ASR) tasks, adversarial training 
on privacy and utility objectives can be used~\cite{AdvRL, gender_adversarial}. Adversarial representation learning determines the network weights that minimise the likelihood of finding the correct labels corresponding to identity (or gender) and removes information related to these attributes from the encoded representation~\cite{AdvRL}.
Training with a speaker-adversarial branch acting as a gradient reversal layer has been used to remove speaker identity from the learned representation~\cite{grl}.
Adversarial feature extraction is used to improve the privacy-utility trade-off, when gender classification is considered as the utility objective and speaker identification accuracy as the privacy objective~\cite{gender_adversarial}.

Gender information is typically used to condition models preserving the identity of a speaker. However, only a handful of methods explicitly consider gender as a sensitive attribute to protect~\cite{Wu, noe, aloufi, pcmelgan, stoidis21_interspeech}.
A hybrid model combining Variational Autoencoders and Generative Adversarial Networks (GANs) can be used to protect gender information through voice conversion with a disentanglement approach targeted for the speech recognition task~\cite{Wu}. Two encoders are trained to independently encode content and speaker identity information that is then used to hide (or mask) gender information. 
Privacy methods that operate at feature-level have been used to disentangle gender information from x-vectors~\cite{snyder} with adversarial learning and an encoder-decoder based architecture~\cite{noe}. Because this adversarial method removes the unwanted information at the level of the feature representation instead of the speech waveform, it is not useful for tasks such as speech recognition.
Vector Quantized Variational Autoencoders (VQ-VAE)~\cite{vq-vae} are used to create disentangled representations of linguistic content and speaker identity~\cite{aloufi}, and both identity and gender~\cite{stoidis21_interspeech}.
Linguistic content is quantised into a discrete latent space using a learned codebook.
The decoding stage reconstructs the speech waveform by combining learned embedding spaces encoding the selected attributes.
These methods have limited reconstruction capabilities and induce distortion at the decoding stage, when quantised content information is reconstructed as speech.
PCMelGAN~\cite{pcmelgan} synthesises speech using a generative adversarial approach that considers gender as an attribute to protect and reconstructs mel-scale spectrograms using the MelGAN vocoder~\cite{melgan} to maintain the utility (accuracy) in a digit recognition task. 
However, the dataset used by this work is composed of utterances
of spoken digits, which is limited in vocabulary and size.
Finally, PCMelGAN is based on PCGAN~\cite{pcgan}, which uses a filtering module to replace the sensitive information in speech with generated synthetic information. However, we will show that results can be improved \emph{without} this additional process.

We aim to produce gender-ambiguous~\cite{genderless} voices.
To this end we first determine the extent to which modifying gender information impacts the identity information of a speaker from a privacy-preservation perspective.
Next, we produce a light-weight generative method that protects against the inference of gender and identity of synthesised speech signals. We achieve this without considering any identity information and maintain the utility of speech without explicitly optimising for the ASR task. 

We propose GenGAN, a generative privacy transformation method that conceals gender and much of the identity information of speakers by synthesising voices with gender-ambiguous characteristics. To the best of our knowledge GenGAN is the first attempt to create gender-ambiguous voices in a privacy-preserving setting.
The generator samples from a designed distribution that models a gender-ambiguous voice and learns to smooth spectral differences between genders.
GenGAN is only conditioned on gender information, without considering any information on speaker identity during training. Furthermore, content information is preserved in the transformed signal independently of the ASR used for the speech transcription task.

\section{Proposed approach}
\label{sec:method}

\subsection{Attack scenario}\label{attack}

Let the identity and gender of a speaker be the personal information to protect in an ASR scenario. An attacker attempts to infer, from the privacy-transformed audio, the gender of the speaker by classification and their identity
by verification.

\begin{figure}
    \centering
    \includegraphics[width=0.47\textwidth]{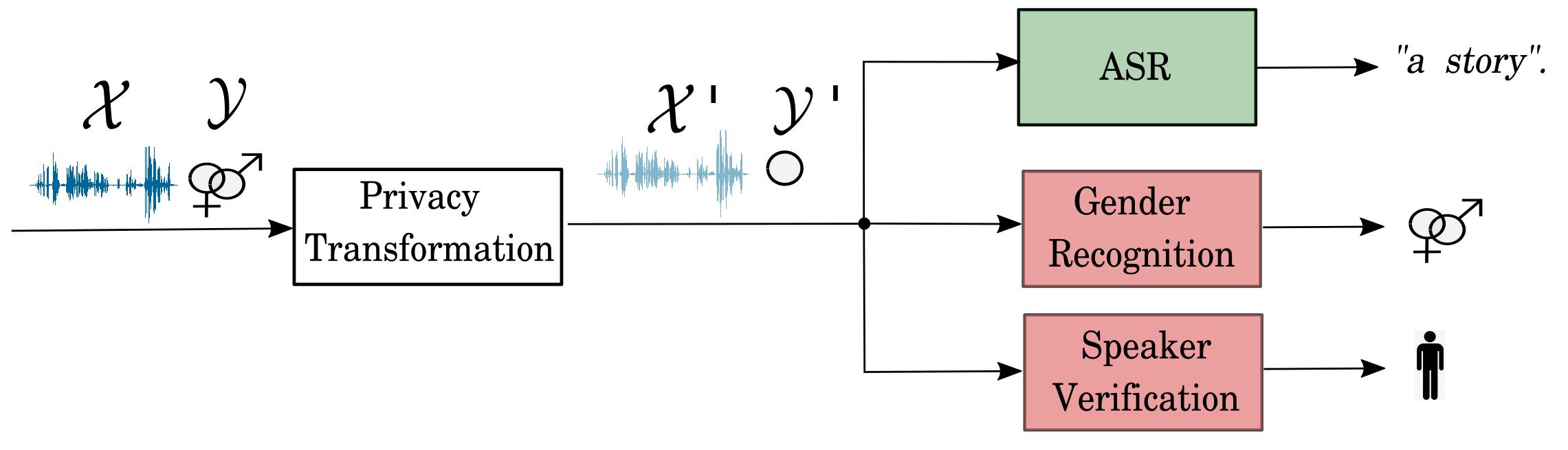}
    \caption{Evaluation of our privacy transformation method on utility \sqbox1{cgreen} and privacy \sqbox1{cred} objectives. The transformed audio is used to perform privacy-preserving speech recognition. The privacy objective assesses the ability of an attacker to recognise the gender and verify the identity of a speaker given the transformed gender-ambiguous speech signal.
    KEY -- $\mathcal{X}$: original waveform, $\mathcal{X'}$: transformed waveform, $\mathcal{Y}$: ground-truth gender, $\mathcal{Y'}$: transformed gender, ASR: automatic speech recognition.}
    \label{fig:attacker}
\end{figure}

We assume that the attacker has access to the data (anonymised utterances) produced by the privacy-preserving transformation and shared with the speech recognition service (see Fig.~\ref{fig:attacker}).
We also assume the attacker has no knowledge of the applied privacy-transformation.

\subsection{GenGAN}

By assuming an attack scenario where voice signals are accessed prior to the downstream task (speech recognition), we are interested in reproducing voice and hence operate on the input waveforms, which are converted into 80-band mel-spectrograms.
We consider a Generator, $\mathcal{G}$, and a Discriminator, $\mathcal{D}$, to be trained in an adversarial game between privacy and utility objectives.
$\mathcal{G}$ has a U-Net architecture~\cite{UNet} with a contracting and an expanding path, while $\mathcal{D}$ consists of a simple AlexNet~\cite{alexnet} architecture.
$\mathcal{G}$ produces audio data with the aim to fool $\mathcal{D}$, whose task is to discriminate between original and synthetic audio data.
We {\em maximise} utility by minimising the distortion in the generated audio and {\em minimise} the risk of privacy inference by training the model to learn gender-ambiguous information. $\mathcal{D}$ learns to discriminate between true and generated gender data, conditioned only on gender information. By conditioning only on gender information, we aim to distort the generated voice to protect both gender and identity.

\begin{figure}[t]
    \centering
    \includegraphics[width=0.47\textwidth]{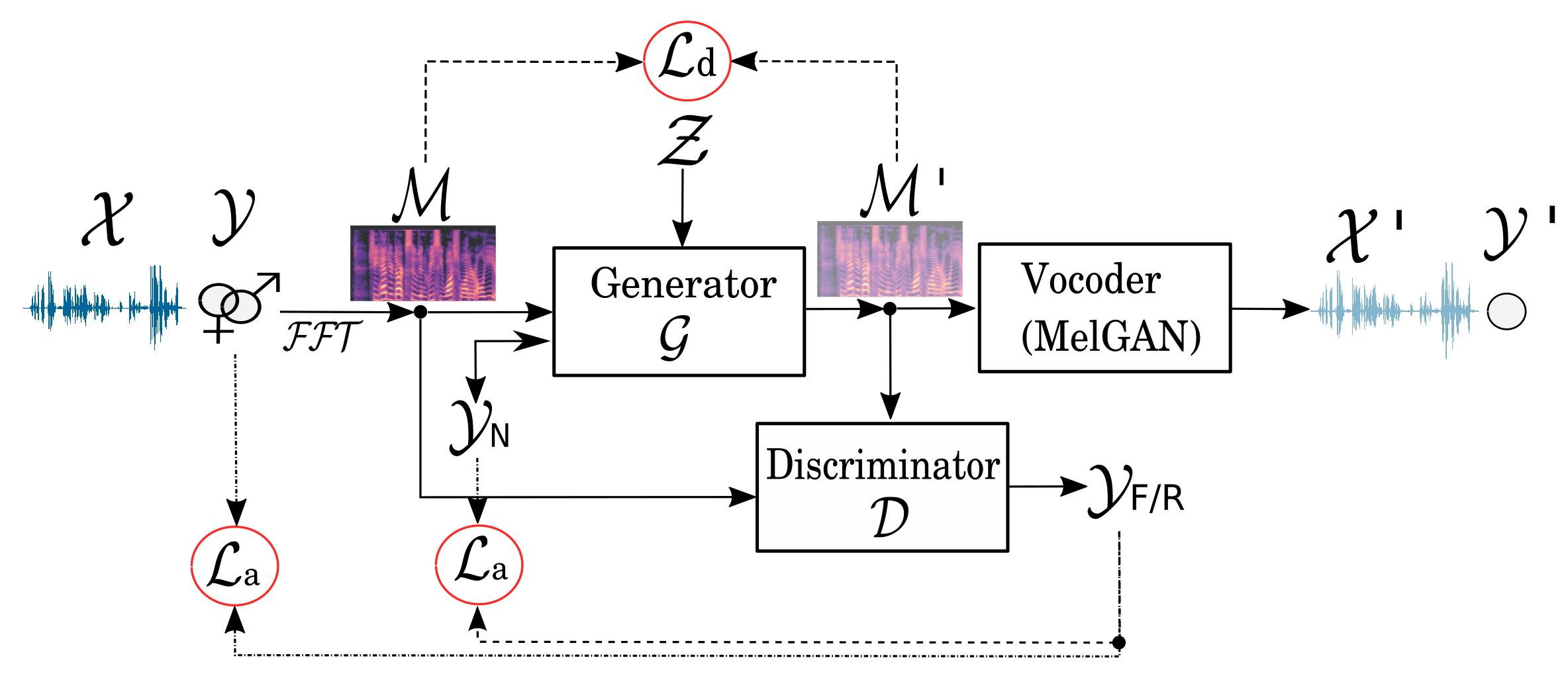}
    \caption{Privacy transformation and training procedure with the losses we use for GenGAN. The time-signal waveform $\mathcal{X}$ is first converted into mel-spectrogram and then fed as input to the Generator along with the noise vector $\mathcal{Z} \sim \mathcal{N}(0,1)$ and the synthetic non-binary gender label $\mathcal{Y}_{N}$. The generated mel-spectrogram $\mathcal{M'}$ is fed to the Discriminator to predict if the sample corresponds to the ground-truth binary gender or non-binary gender. We use the MelGAN pre-trained vocoder to perform the mel-spectrogram inversion. KEY -- $\mathcal{X}$: original waveform, $\mathcal{X'}$: transformed waveform, $\mathcal{M}$: original mel-spectrogram, $\mathcal{M'}$: transformed mel-spectrogram,
    $\mathcal{Y}$(\FemaleMale): ground-truth gender label,
    $\mathcal{Y}_{N}$: non-binary gender vector, $\mathcal{Y'}$(\Neutral): synthesised gender, $\mathcal{Y}_{F/R}$: predicted gender with Discriminator from real (original) and fake (generated) spectrograms, 
    $\mathcal{L}_d$: distortion loss,
    $\mathcal{L}_a$: adversarial loss,
    $\mathcal{FFT}$: Fast Fourier Transform.}
    \label{fig:model_architecture}
\end{figure}

The Generator, $\mathcal{G}$, takes as input mel-spectrograms $\mathcal{M}$, a noise vector $\mathcal{Z}$ and the labels of the sensitive attribute to protect $\mathcal{Y}$, and synthesises the transformed spectrograms $\mathcal{M}'$ (see Fig.~\ref{fig:model_architecture}).
During training, a batch of $n$ audio signals, $\mathcal{X}$, and their corresponding labels, $\mathcal{Y}$, representing the sensitive attribute (gender) are sampled uniformly at random from the dataset:
\begin{equation}
(x_1,y_1),...,(x_n,y_n) \sim \mathcal{X}_{train}. 
\end{equation}

The audio signals $\mathcal{X}$ are converted to mel-spectrograms $m_i \in \mathcal{M}$ $\forall i=1,...,n$, and normalized such that amplitudes are bounded in $[0,1]$ with
\begin{equation}
m_i = \mathcal{FFT}(x_i), \quad  
\end{equation}
where $\mathcal{FFT}$ is the Short-Time Fourier Transform.

As we consider a binary encoding for gender, we propose to sample from a synthetic distribution 
\begin{equation}
    \{\hat{y}_1,...,\hat{y}_n\} \in \mathcal{Y}_{N}\sim \mathcal{N}(0.5, 0.05).
\end{equation}

As $\mathcal{N}(0.5, 0.05)$ is centred around 0.5 (i.e.~equidistant from the ground-truth labels), $\mathcal{G}$ learns to smooth feature-level differences between the spectral representations of the two gender labels and synthesises a new voice that is gender-ambiguous. We select a small distribution variance ($\sigma^2 = 0.05$) for the synthetic distribution to minimise the overlap with the gender distributions. The noise vector $\mathcal{Z}\sim \mathcal{N}(0,1)$ is inserted at the bottleneck of $\mathcal{G}$, at the transition of the contracting to the expansive path to ensure synthesised voices are different from the original ones, increasing reconstruction variability.
$\mathcal{Z}$ is reshaped and concatenated with the last convolutional layer ending the contracting path in UNet before being expanded back to a mel-spectrogram $\mathcal{M}'$.

The generator loss $\mathcal{L}_{\mathcal{G}}$ is computed as an adversarial game between reducing distortion on utility and maximising privacy by learning a non-binary gender.
We take the Mean Squared Error between $\mathcal{M}$ and $\mathcal{M}'$ as distortion loss $\mathcal{L}_{d}$ over the $L_1$ distance used in PCMelGAN to produce a smoother output.
The adversarial loss $\mathcal{L}_{a}$ is a cross-entropy loss between ground-truth gender $\mathcal{Y}$ and predicted gender $\mathcal{Y}_{F}$, maximising the log probability of generating samples drawn from the synthetic distribution. The generator loss is the sum

\begin{equation}\label{gen_eq}
    \mathcal{L}_{\mathcal{G}} = \mathcal{L}_{d}(\mathcal{M},\mathcal{M')} +\epsilon \mathcal{L}_{a}(\mathcal{Y},\mathcal{Y}_{F}),     
\end{equation}
where $\epsilon \in [0,1]$ represents the dis-utility budget~\cite{tripathy2019privacyadvnn} in the privacy-utility trade-off.

The discriminator loss function is composed of two losses with respect to real $\mathcal{M}$ or generated spectrogram $\mathcal{M}'$.
The real loss is taken between the prediction of $\mathcal{D}$ on real data $\mathcal{Y}_{R}$, and ground-truth gender $\mathcal{Y}$.
The fake loss is taken between the prediction of $\mathcal{D}$ on generated data $\mathcal{Y}_{F}$, and non-binary gender $\mathcal{Y}_{N}$.
Both losses are computed as cross-entropies:
\begin{equation}\label{disc_eq}
    \mathcal{L}_{\mathcal{D}} = \mathcal{L}_{a}(\mathcal{Y},\mathcal{Y}_{R}) + \mathcal{L}_{a}(\mathcal{Y}_{N}, \mathcal{Y}_{F}).
\end{equation}

Finally, to transform the generated spectrogram $\mathcal{M}'$ back to raw waveform $\mathcal{X}'$, we use the MelGAN vocoder~\cite{melgan}, a non-autoregressive conditional waveform synthesis model.

\begin{figure}[t]

     \begin{subfigure}[b]{0.22\textwidth}
         \centering
         $\quad$\text{Male}
         \includegraphics[width=\textwidth]{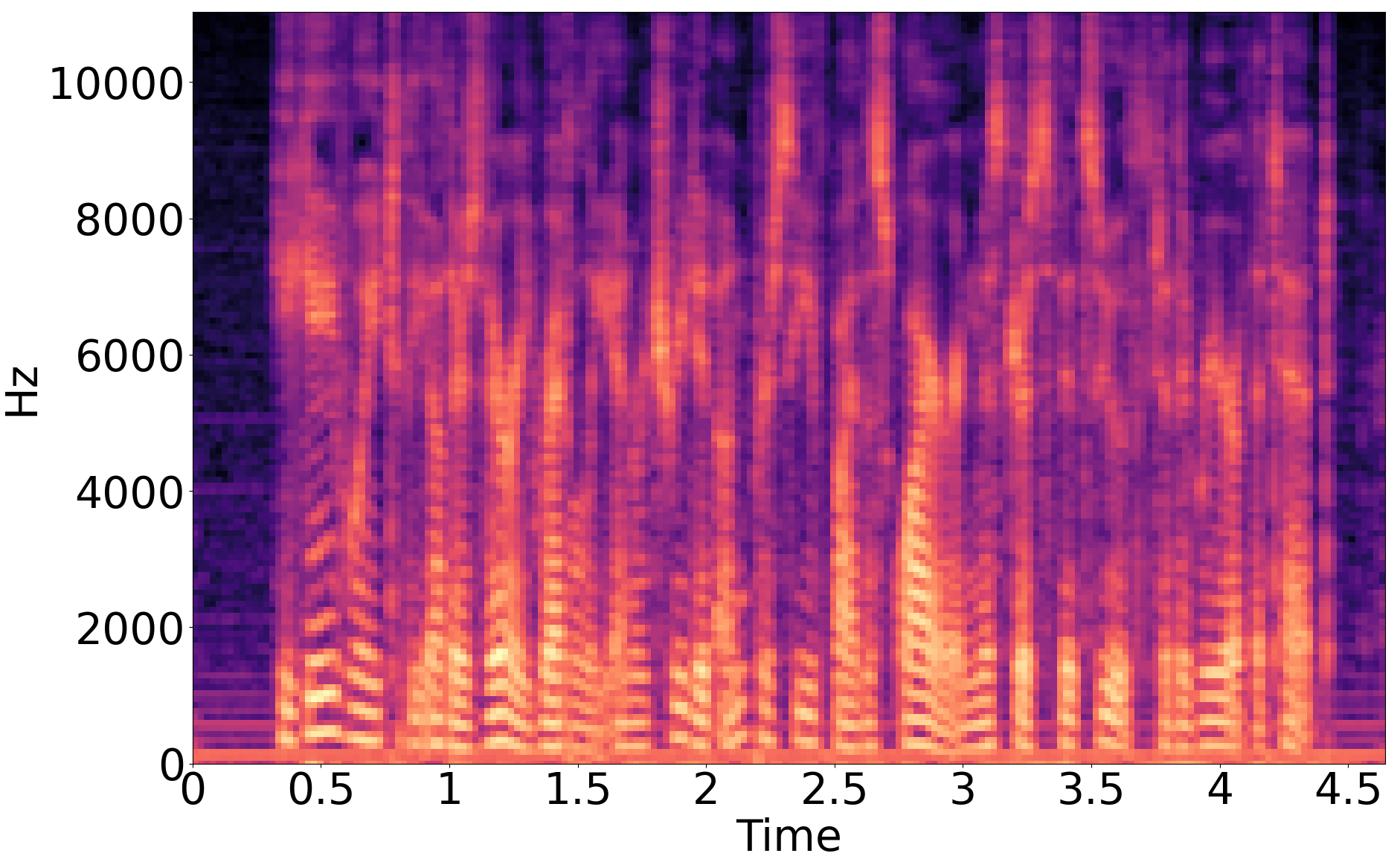}
         \caption{Original}
         \label{fig:orig_spec}
     \end{subfigure}
      \hfil
     \begin{subfigure}[b]{0.22\textwidth}
         \centering
         $\quad$\text{Female}
         \includegraphics[width=\textwidth]{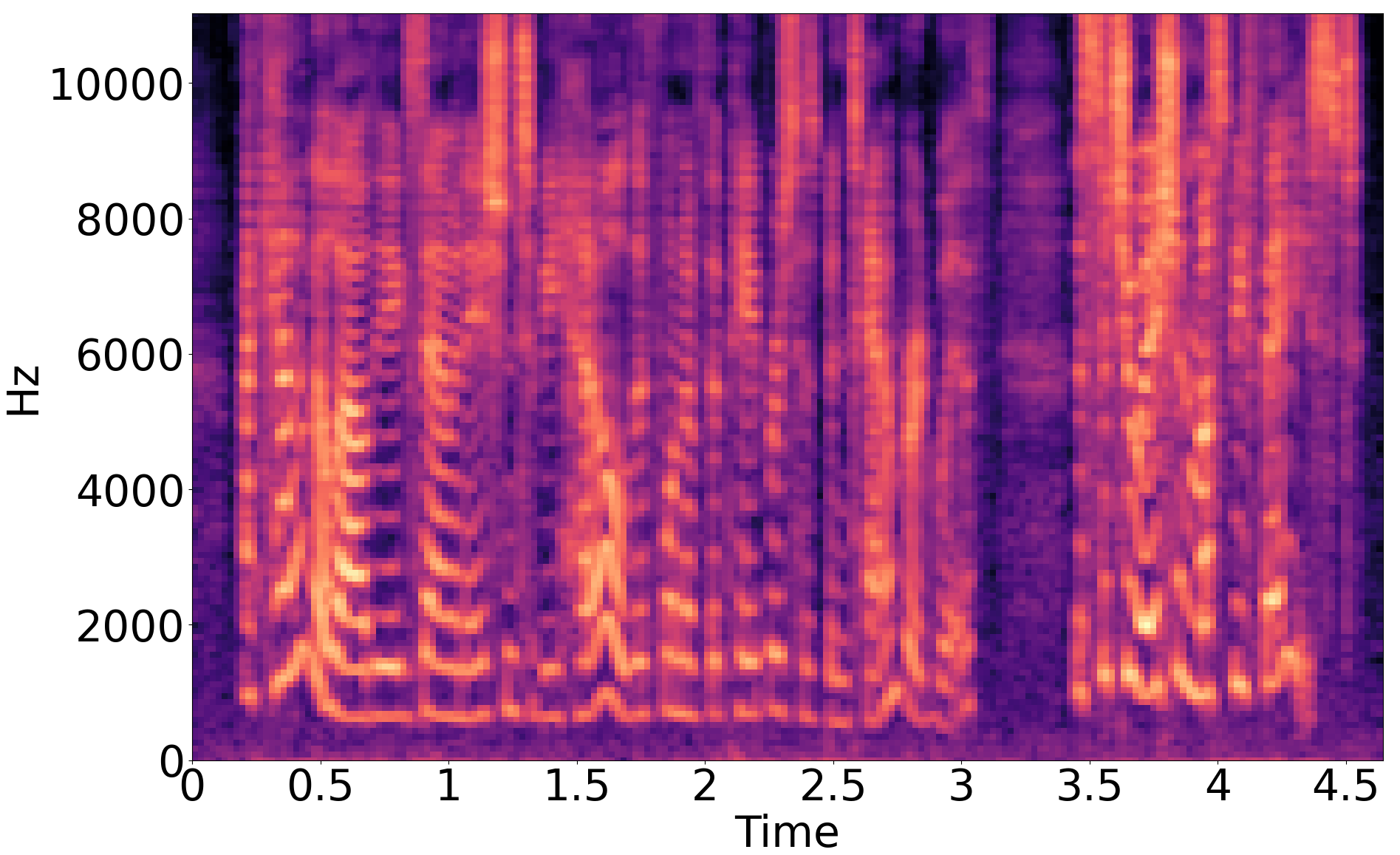}
         \caption{Original}
         \label{fig:orig_specf}
     \end{subfigure}
     \hfil
      \begin{subfigure}[b]{0.22\textwidth}
         \centering
         \includegraphics[width=\textwidth]{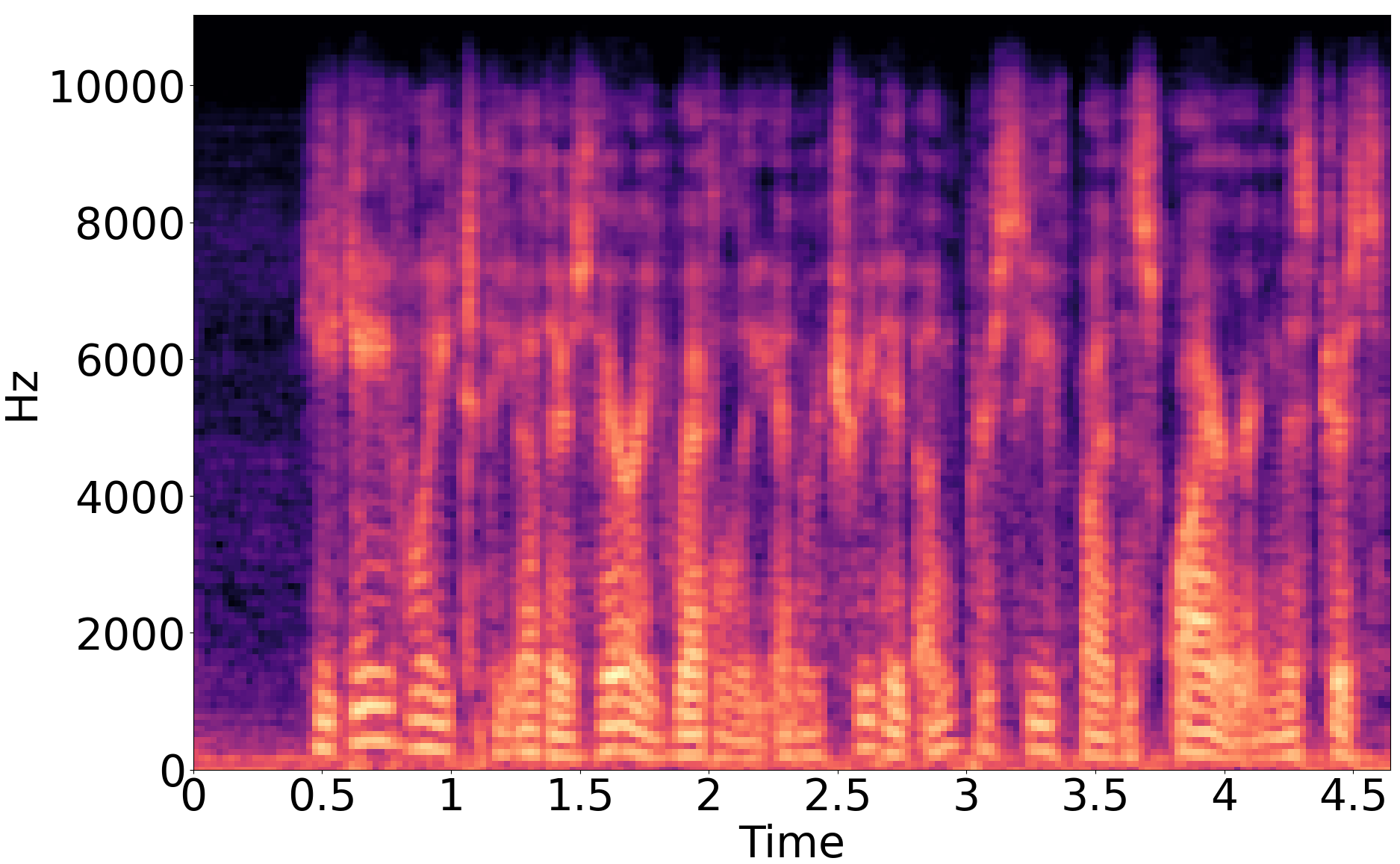}
         \caption{MelGAN}
         \label{fig:melgan_spec}
     \end{subfigure}
     \hfil
      \begin{subfigure}[b]{0.22\textwidth}
         \centering
         \includegraphics[width=\textwidth]{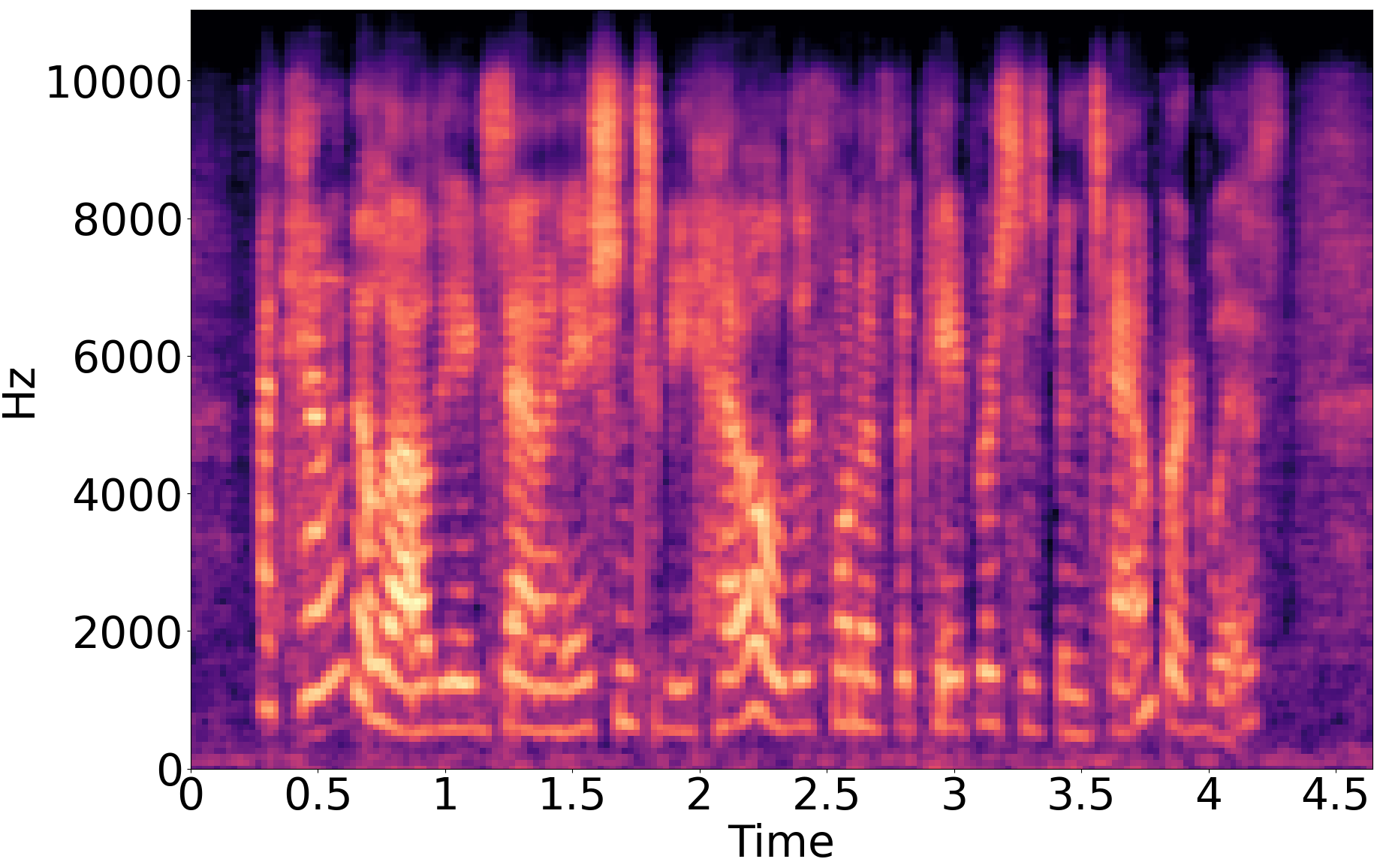}
         \caption{MelGAN}
         \label{fig:melgan_specf}
     \end{subfigure}
    \hfil
     \begin{subfigure}[b]{0.22\textwidth}
         \centering
         \includegraphics[width=\textwidth]{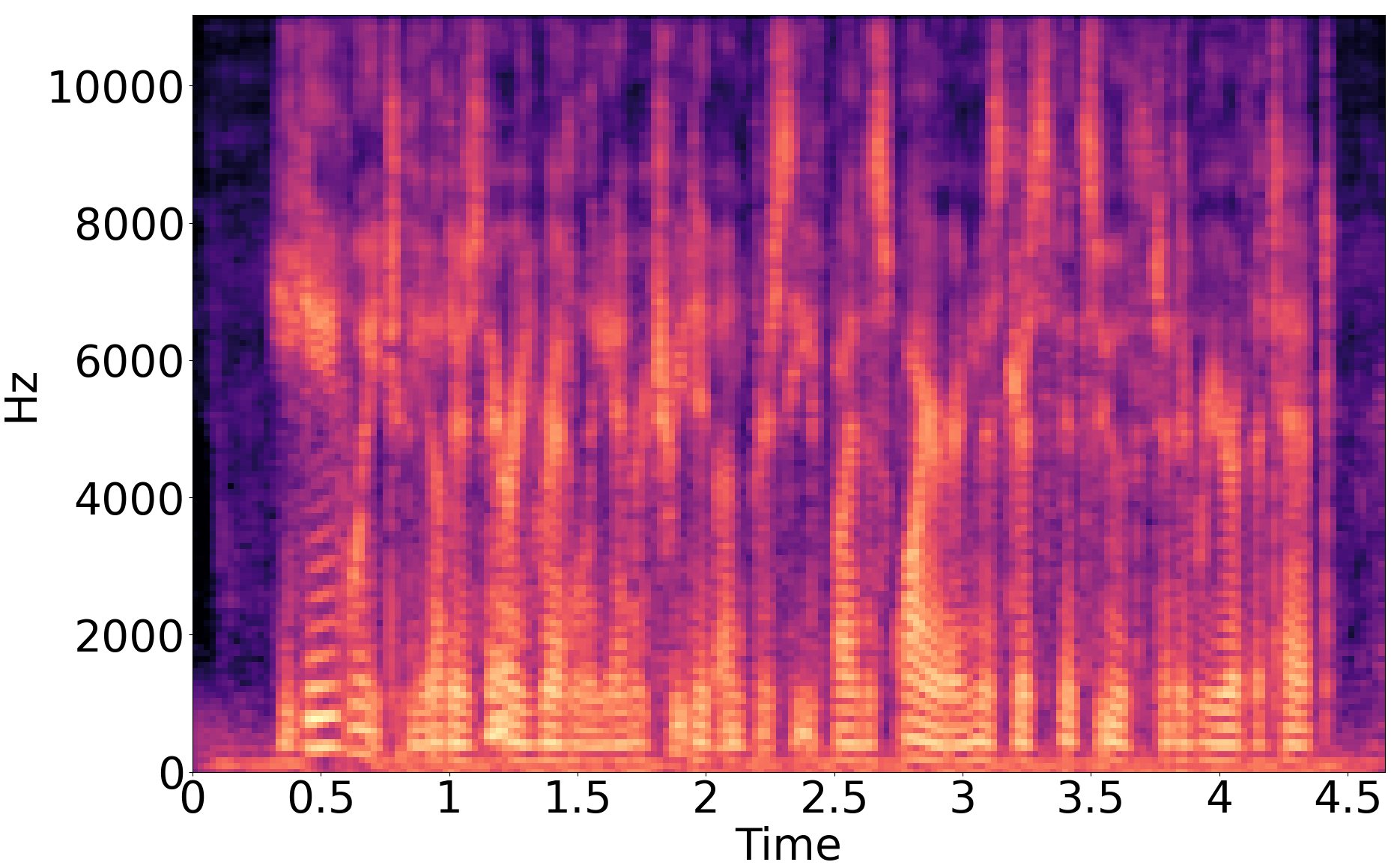}
         \caption{PCMelGAN}
         \label{fig:pcmelgan_spec}
     \end{subfigure}
     \hfil
     \begin{subfigure}[b]{0.22\textwidth}
         \centering
         \includegraphics[width=\textwidth]{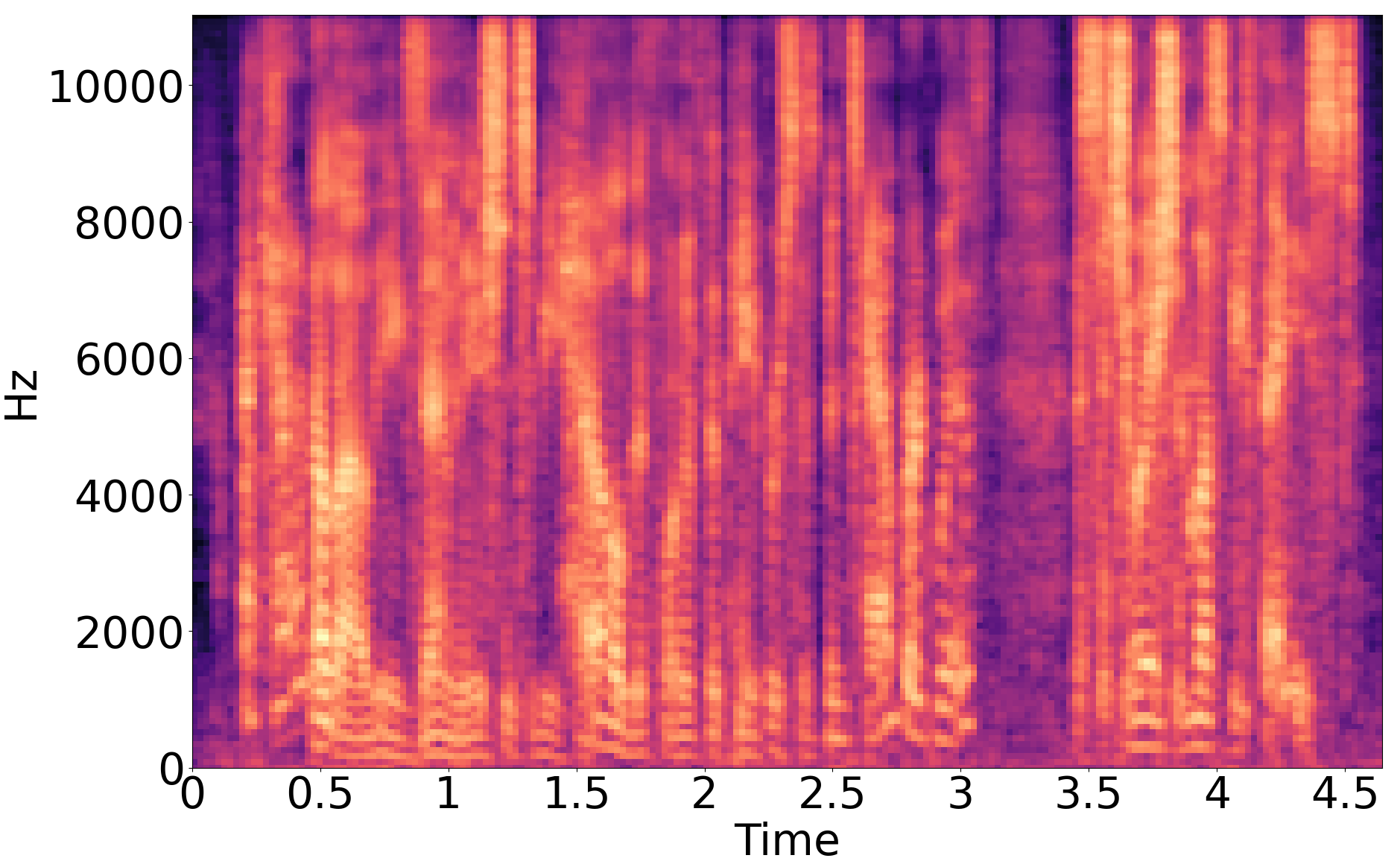}
         \caption{PCMelGAN}
         \label{fig:pcmelgan_specf}
     \end{subfigure}
    \hfil
     \begin{subfigure}[b]{0.22\textwidth}
         \centering
         \includegraphics[width=\textwidth]{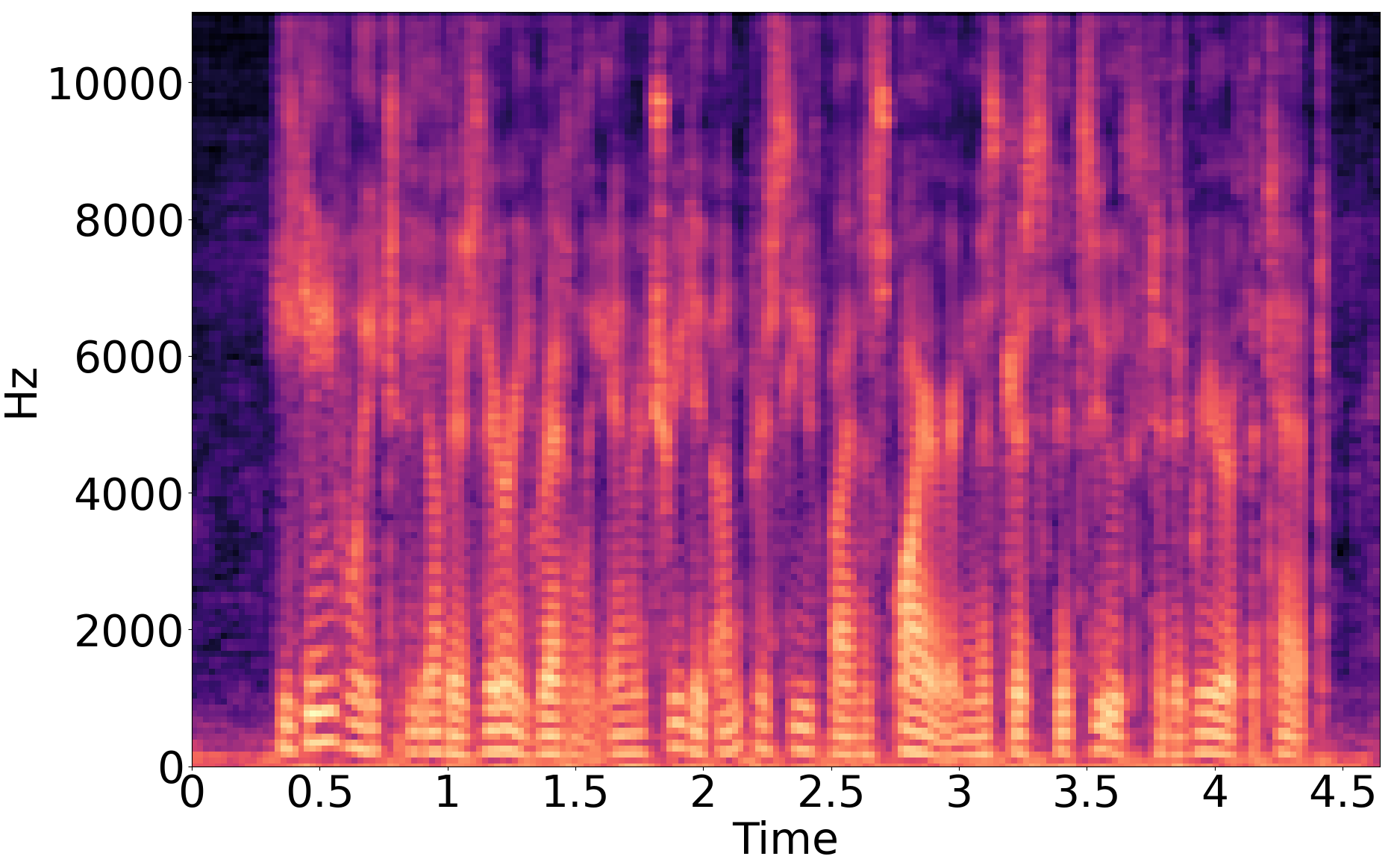}
         \caption{GenGAN}
         \label{fig:gengan_spec}
     \end{subfigure}
     \hfill
     \begin{subfigure}[b]{0.22\textwidth}
         \centering
         \includegraphics[width=\textwidth]{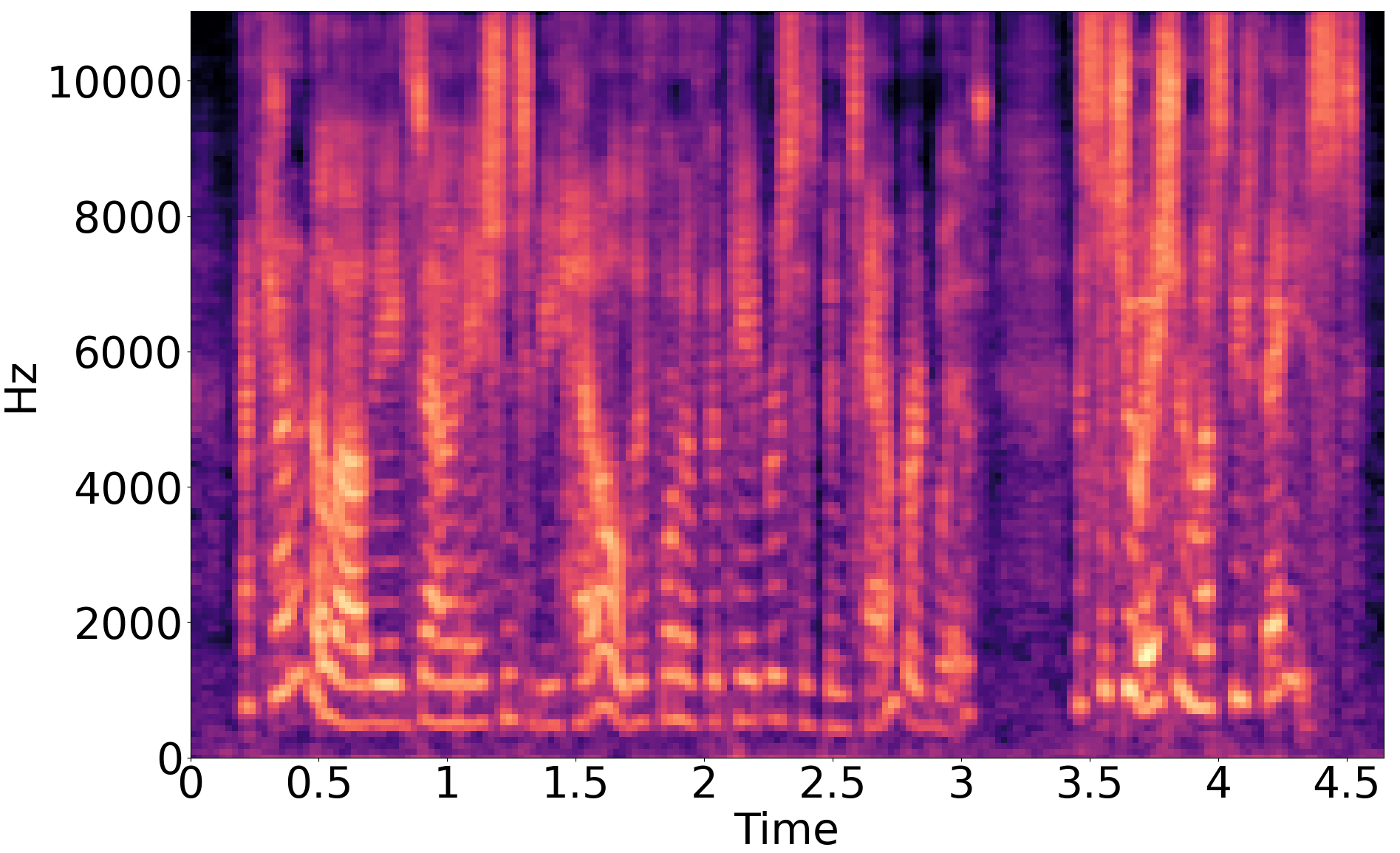}
         \caption{GenGAN}
         \label{fig:gengan_specf}
     \end{subfigure}

        \caption{Spectrograms of two different utterances spoken by a male (left column) and a female speaker (right column). 
        The privacy-preserving transformation affects the high formant frequencies associated with gender (see (e), (f), (g), (h)).
        (c) and (d) show the distortion induced on high frequencies by the MelGAN vocoder with respect to the original utterances.}
        \label{fig:specs}
\end{figure}

Figure~\ref{fig:specs} shows sample spectrograms of different utterances spoken by a male and by a female speaker. The bottom row shows the privacy-transformed spectrograms produced by GenGAN. 
GenGAN synthesises higher resolution spectrograms than PCMelGAN~\cite{pcmelgan}, while both methods affect high formant frequencies associated with gender identification~\cite{formants}, smoothing out differences between male and female voices.

\section{Validation}

 
In this section we perform the validation of GenGAN in comparison with existing attribute-protecting privacy-preserving methods. 
We use the LibriSpeech dataset~\cite{librispeech}, which contains over 100 hours of transcribed clean speech and over 250 different speakers with balanced gender distribution.

\subsection{Performance measures}

We now describe the measures we use to compare the performance of different models in the scenario presented in Section~\ref{attack}.

We view speaker {\em identity} in the context of a speaker verification task, which consists in determining whether a trial utterance corresponds to the utterance of a certain, enrolled speaker.
We chose speaker verification instead of speaker identification as in our scenario the attacker uses a speaker verification system to recover the identity of a speaker from anonymised speech.
We consider the equal error rate (EER) to assess the speaker verification performance. 
In biometrics verification systems, a high accuracy corresponds to a low EER, as a higher rejection over false acceptance rate is desired. To assess privacy, the target randomness in verifying the speaker identity~\cite{tomashenko2020introducing} corresponds to an EER around 50\%.

We measure success in {\em gender} inference as the standard binary classification accuracy, where the sum of correct predictions is taken over the total number of predictions.
As in speaker verification, the target accuracy is 50\%, which represents randomness in gender prediction.
We compute the discrepancy from the 50\% randomness value both on gender recognition and speaker verification and introduce a new measure.
The measure normalises the EER and the gender recognition accuracy (GR) such that the absolute difference from the 50\% accuracy is retained.
The normalised Gender Recognition ($gr$) and normalised Equal Error Rate ($eer$) perform conversions given by $gr = 100 -2\times \lvert GR-50\rvert$, and $eer= 100 -2\times \lvert EER-50 \rvert$.
A value of 100 for $gr$ (or $eer$) denotes the highest level of privacy.

We use the transcription accuracy of the speech recognition results as the utility measure.
The Word Error Rate (WER) represents the utility of the spoken signal and
is computed by taking the Levenshtein~\cite{levenshtein} distance between the words of the correct, expected transcription and the transcription provided by the speech recognition service. 
To facilitate the comparison between privacy and utility, rather than the error rate we consider the correct recognition rate with the use of the Word Accuracy metric ($A_w$)~\cite{stoidis21_interspeech, hidebehind}, where ${A_w} = 100 - WER$. $A_w = 100$ denotes no transcription errors.

\subsection{Classifiers}
For {\em speech recognition}, we use Quartznet~\cite{quartznet}, an end-to-end neural acoustic model trained with Connectionist Temporal Classification (CTC)~\cite{ctc} loss, based on the Jasper architecture~\cite{jasper}.
Our testing reported an initial performance on the LibriSpeech test-clean set of $A_w$ of 95.64\% (or 4.36\% WER).

For {\em speaker verification}, we extract speaker-identifying features with a modified version of ResNet-34~\cite{Resnet34} with reduced computational cost, which we refer to as SpeakerNet~\cite{speakernet}, pre-trained on VoxCeleb2~\cite{voxceleb2} dataset and reported an EER of 5.73\% in our experiments when tested on the LibriSpeech test-clean set.

For {\em gender classification}, we use a deep convolutional binary classifier trained on spectrograms with 5 stacked one-dimensional convolutional layers followed by batch normalisation and max pooling~\cite{stoidis21_interspeech}, which we refer to as GenderNet.
The output layer is a fully connected layer that outputs a pair of predictions for each binary class, passing through a sigmoid function.
We tested the gender classifier on the LibriSpeech clean test set and reported an accuracy of 91.37\%.

\subsection{Methods under comparison}

We compare methods that consider gender as an attribute to protect in an attribute inference scenario.
VQ-VAE-Gen~\cite{stoidis21_interspeech} considers gender information as private and assesses the impact of the privacy-transformation on speaker identity on LibriSpeech.
The Client-Hybrid~\cite{Wu} model protects gender information from inference by using speaker identity information and provides results on the same dataset.
We also compare our method with PCMelGAN~\cite{pcmelgan} anonymisation method which uses a filtering process to remove gender information and conditions the model with identity information during training. We did not modify the architecture of PCMelGAN and trained it from scratch on the LibriSpeech clean train set.
We improve and address the limitations in PCMelGAN's pipeline with GenGAN's implementation. We train PCMelGAN with the additional Filter network prior to the generation and keep the original loss function used in~\cite{pcmelgan} for the same number of epochs and identical hyper-parameter setting.
We use the same pre-trained MelGAN vocoder model in our experiments with PCMelGAN and GenGAN.
We also use the same pre-trained models to evaluate the privacy tasks for GenGAN, PCMelGAN and VQ-VAE-Gen. For utility, we use the same ASR for GenGAN and PCMelGAN models for a fair comparison.
We run all our experiments on a single Tesla V100 GPU with 32GB memory. 
The models were trained for 100 epochs by shuffling batches, a learning rate of 0.001 and Adam optimiser~\cite{adam}.

\subsection{Discussion}

\begin{table}[t]
    \footnotesize
    \caption{Comparison of privacy and utility results for various models on the Librispeech~\cite{librispeech} test clean set. KEY -- $A_w$: Word Accuracy computed with QuartzNet~\cite{quartznet}, EER: Equal Error Rate computed with SpeakerNet~\cite{speakernet}, $eer$: normalised EER,
    GR: Gender Recognition computed with GenderNet~\cite{stoidis21_interspeech}, $gr$: normalised Gender Recognition,
    $\dagger$: RG (Random Gender) setting~\cite{stoidis21_interspeech} is reported. $\star$: a random accuracy close to 50\% is desired for high privacy. 
    Results for GenGAN are obtained with $\epsilon= 0.001$ and $\mathcal{Y}_N\sim (0.5, 0.05)$.
    The MelGAN model is reported along with the Original signal to assess the impact of the spectrogram inversion without privacy-preserving transformation.}
    
    \centering
    \begin{tabular}{l|c|cccc}
        \hline
        \textbf{Model}  &\multicolumn{1}{c|}{\textbf{Utility}} &\multicolumn{4}{c}{\textbf{Privacy}} \\
        & $A_w$ $\uparrow$ & EER$\star$ & $eer$ $\uparrow$ & GR$\star$ & $gr$ $\uparrow$\\
        \hline
         Original   & 95.64  & 5.73 & 11.46 & 91.37 & 17.26\\
        {MelGAN~\cite{melgan}} & 93.22 & 17.42 & 34.84 & 89.04 & 21.92\\
        \hline
          Client-Hybrid~\cite{Wu}  & 71.54 & \textendash & \textendash & 53.90 & 92.20 \\
          {VQ-VAE-Gen~\cite{stoidis21_interspeech}}$\dagger$ & 26.84 & 51.88 & 96.24 & 50.01 & 99.98\\
          PCMelGAN~\cite{pcmelgan} & 66.81 & 41.95 & 83.90 & 48.39 & 96.78\\
          GenGAN (ours) & 76.64 & 38.37 & 76.74 & 53.63 & 92.74 \\
          \hline
    \end{tabular}
    \label{tab:exper_all4}
\end{table}

In Tab.~\ref{tab:exper_all4} ``Original'' refers to the original audio prior to any transformation.
We also include the MelGAN~\cite{melgan} vocoder model to assess the distortion induced by the mel-spectrogram inversion and compare the effect of this operation on utility and privacy of the generated signal.

The values of privacy and utility for Client-Hybrid~\cite{Wu} and VQ-VAE-Gen~\cite{stoidis21_interspeech} are taken from the literature. The values corresponding to the `Hybrid' model with fine-tuning on two speakers were chosen for Client-Hybrid, and the `Random Gender' (RG) setting, which considers only gender information during training, for VQ-VAE-Gen. 
Although VQ-VAE-Gen provides the best privacy guarantees both on gender and identity inference, it is also the worst performing in terms of utility with an $A_w$ value of 26.84\%. 
The Client-Hybrid model performs equally well in terms of utility and privacy, however the selected model for comparison was fine-tuned on two speakers from the train set while no results on the speaker verification task are provided.
Results for PCMelGAN show that the privacy objectives are achieved but utility remains low ($A_w=66.81$).
GenGAN improves upon the utility with respect to existing related models without being trained explicitly to optimise for the speech recognition task.
GenGAN is a simplified version of PCMelGAN with modified cost functions that improve the privacy-utility trade-off with respect to methods considering gender information.
In terms of privacy, the 76.74 $eer$ value suggests that some residual identity information is still present in the generated speech, since identity information was not explicitly manipulated in the GenGAN pipeline. 

\begin{figure}[t]
    \centering
      \includegraphics[scale=0.135]{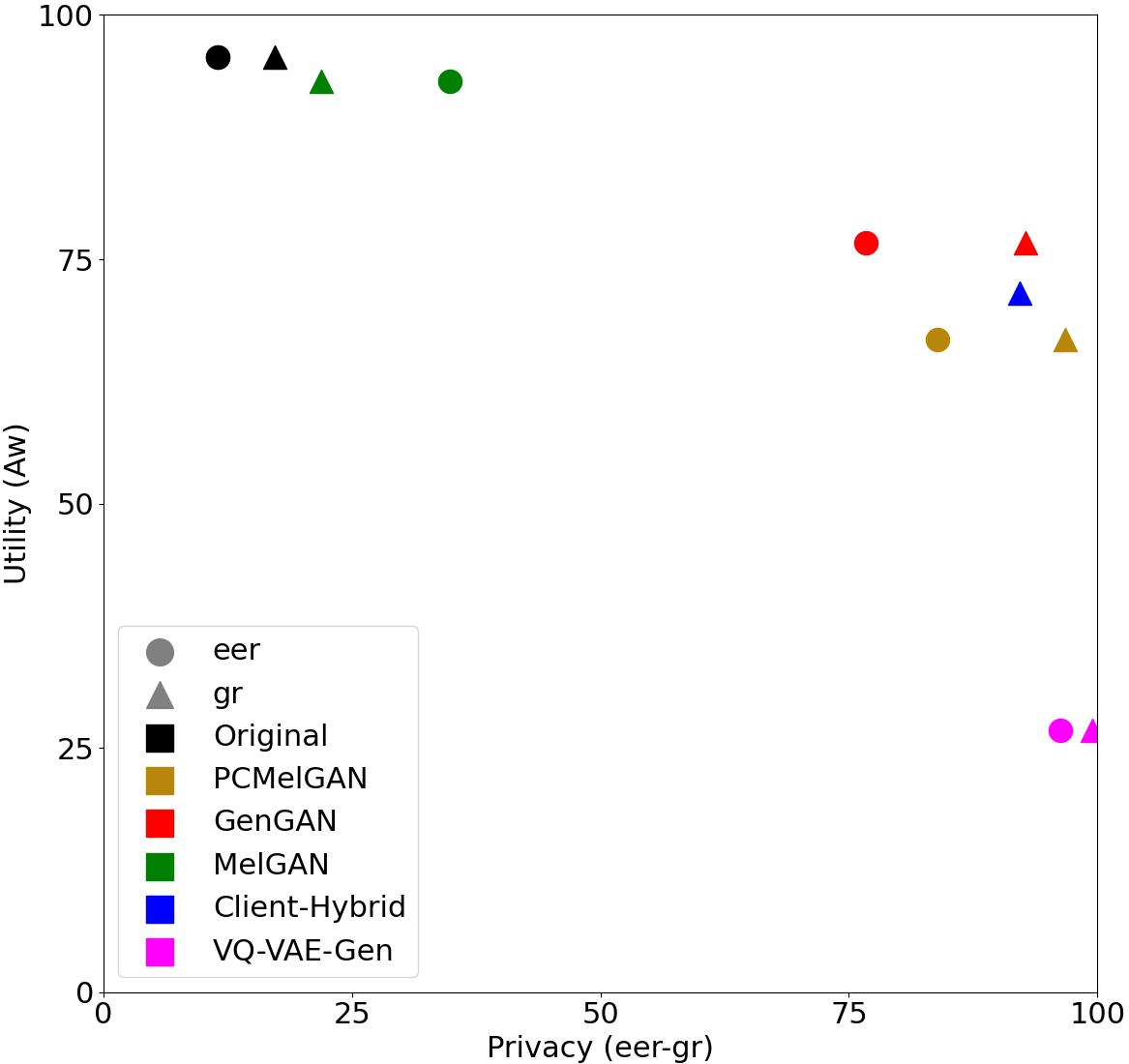}
     \label{fig:pu_nGR}
    \caption{Comparisons of the privacy-utility trade-off for the methods compared in Table 1. Values close to the top-right corner denote higher utility and privacy. GenGAN provides the best utility and comparable privacy performance. KEY -- $A_w$: Word Accuracy, $gr$: normalised Gender Recognition, $eer$: normalised Equal Error Rate.}
    \label{fig:p_utradeoff}
\end{figure}

Privacy-utility results are reported in Fig.~\ref{fig:p_utradeoff}. The larger $eer$, $gr$ and $A_w$, the higher the privacy and the higher the utility.
GenGAN reaches the best trade-off with improved reconstruction capabilities and comparable privacy performance.

\section{Conclusion}
\label{sec:conclusion}

We proposed GenGAN, a generative adversarial network that synthesises gender-ambiguous voices that can be used for privacy-preserving speech recognition scenarios.
The generator and discriminator are adversarially trained to limit the distortion in the resulting gender-ambiguous signal.
Our model improves the privacy-utility trade-off with respect to existing methods.

Future work includes improving the signal reconstruction
capabilities of the network without compromising privacy and assessing the naturalness of the generated voices with subjective evaluations.

\bibliographystyle{IEEEtran}

\bibliography{mybib}

\end{document}